\documentclass[conference]{IEEEtran}
\IEEEoverridecommandlockouts
\usepackage{cite}
\usepackage{amsmath,amssymb,amsfonts}
\usepackage{verbatim}
\usepackage{algorithmic}
\usepackage[T1]{fontenc}
\usepackage{float}
\usepackage{fancyhdr}
\usepackage{caption}
\usepackage{graphicx}
\usepackage{url}
\usepackage{gensymb}
\usepackage{xcolor}
\usepackage{hyperref}

\newcommand{\mycomment}[1]{}
\usepackage{textcomp}

\def\BibTeX{{\rm B\kern-.05em{\sc i\kern-.025em b}\kern-.08em
    T\kern-.1667em\lower.7ex\hbox{E}\kern-.125emX}}

\title{{\large 2025 IEEE International Students' Conference on Electrical, Electronics, and Computer Science} \\

LifeSaver: Predictive Load Limit Estimation for Transport Vehicles in Hilly Areas

{
\thanks{979-8-3315-2983-3/25/\$31.00 ©2025 IEEE
}
}}

\author{\IEEEauthorblockN{Vaibhav Chopra}
\IEEEauthorblockA{\textit{Computer Science \& Artificial Intelligence} \\
\textit{Plaksha University}\\
Mohali, India \\
0009-0004-9571-757X}
\and
\IEEEauthorblockN{Chanakya Rao}
\IEEEauthorblockA{\textit{Data Science, Economics \& Business} \\
\textit{Plaksha University}\\
Mohali, India \\
0009-0004-2149-0720}
\and
\IEEEauthorblockN{Moksh Soni}
\IEEEauthorblockA{\textit{Robotics \& Cyber-Physical Systems} \\
\textit{Plaksha University}\\
Mohali, India \\
0009-0004-1088-7482}
\and
\IEEEauthorblockN{\hspace{1cm}}
\IEEEauthorblockA{ \hspace{1cm}\\
\hspace{1cm}\\
\hspace{1cm} \\
}
\and
\IEEEauthorblockN{Prashant Mishra}
\IEEEauthorblockA{\textit{Computer Science \& Artificial Intelligence} \\
\textit{Plaksha University}\\
Mohali, India \\
0009-0009-2365-7279}
}

\fancypagestyle{customfooter}{
    \fancyhf{} 
    \fancyfoot[C]{\small{SCEECS 2025} 
}}

\begin{document}

\maketitle

\pagestyle{customfooter}

\begin{abstract}
The transportation of essential goods in mountainous regions faces severe logistical challenges and frequent disruptions. To mitigate these difficulties, transport companies often overload trucks, which, though cost-saving, significantly heightens the risk of accidents and mechanical failures. This paper presents the development of a device that detects overloaded and insecurely fastened loads on trucks and commercial vehicles. Using advanced load sensors, the device offers real-time monitoring of cargo weight distribution, alerting drivers and authorities to unsafe conditions. The initial prototype utilised two basic load cells and an Arduino microcontroller. The second version was enhanced with four load cells and extended sensors. This version was tested by placing an electric golf cart onto the prototype. Various loads were then added to the cart in different orientations to assess whether the system could accurately detect improper or excessive load conditions.

\end{abstract}


\section{Motivations and Objectives}

Mountainous regions present significant transportation challenges due to their complex topography and unpredictable meteorological conditions. The narrow and often precarious roadways in these areas are frequently subjected to disruptions caused by snow accumulation, landslides, and seasonal road collapses. These environmental factors complicate the logistics involved in delivering essential goods such as food, medical supplies, and fuel. To mitigate these challenges and reduce operational costs, transport companies are often incentivised to overload trucks, aiming to minimise the number of trips required for transporting goods [1]. While this approach might appear to be economically advantageous, it introduces substantial stress on vehicle components, including braking systems, suspension assemblies, and tyres. 

\vspace{1\baselineskip}

This excessive load adversely affects vehicle stability and handling, thereby increasing the likelihood of brake failure, vehicle rollovers, and other forms of instability. Apart from this, Aggarwal and Parameswaran [2] have shown how increased load on vehicles causes undue and excess stress on  bridges, leading to deterioration before planned upgrades. Their research shows that ``the increase in truck weight of 50\% would lead to an increase in fatigue damage accumulation of 80\% in the steel longitudinal girder of road bridges". Himachal Pradesh is home to some of the world's finest feats of engineering, including the Chicham bridge, the highest bridge in Asia [3].
In addition, several dams and bridges, such as the Chamera Dam and the Bhakra Nangal Dam, have been developed with a focus on boosting tourism. These locations are designed to be tourist-friendly to maximise tourism revenue. However, the influx of tourists brings with it increased demand for essentials such as food, bottled water, clothing, souvenirs, and fuel. Since railway stations or civilian airports in these regions are few-to-none, all supplies must be transported by road. This logistical pressure, particularly during peak tourist season, incentivises drivers to overload their vehicles to meet the rising demand.

In 2022, trucks, buses, and other commercial vehicles were responsible for approximately 23\% of all vehicular accidents and nearly 19\% of vehicular fatalities in Himachal Pradesh (HP) [4]. In contrast, in Karnataka—a state characterised by relatively flatter terrain—commercial vehicles accounted for only 11.6\% of vehicular accidents and 10.33\% of accident-related deaths [5]. The figures for the hilly state of Himachal Pradesh are nearly double, despite it performing better on several development indices, as measured by NITI Aayog. For instance, during the same period, NITI Aayog assigned an overall score of 63.1 to HP, compared to Karnataka’s score of 61.77. Specifically, HP outperformed Karnataka in governance, scoring 76.76 against Karnataka’s 48.04, and in health systems and delivery, with scores of 55.83 for HP and 52.75 for Karnataka [6],[7]. Therefore, it is unlikely that governance or healthcare infrastructure is the primary factor contributing to the disproportionate rate of accidents in hilly regions. The authors attribute this discrepancy to the prevalence of incorrect and excessive loading of commercial vehicles compounded by the treacherous mountainous areas.

Moreover, the practice of overloading compromises vehicle safety further when cargo is not securely fastened. Insecurely fastened loads have the potential to shift during transit, exacerbating vehicle instability and increasing the risk of accidents. 

The existing regulatory framework also does not fully address the risks associated with overloading, allowing unsafe practices to persist unchecked. 
For example, Sandeep Malhotra, a taxi driver from Nainital, faced the perplexing situation of being denied an insurance claim after his car plunged 200 metres into a gorge. Seeking redressal, he approached the Uttarakhand State Commission of Consumer Grievance Redressal to uphold his consumer rights. However, the commission upheld a lower court’s decision, which stated that an `Additional Endorsement to Driving License' is required for all commercial vehicles operating in hilly regions [8].

At first glance, this ruling may seem rational. In private vehicles, the driver bears the responsibility of ensuring the safe passage of all occupants, a role based on mutual trust and consensus. However, for taxi drivers, the contractual nature of their service imposes an additional fiduciary duty to act in the best interest of their passengers. This should include obtaining the necessary qualifications to drive in hilly areas, where insufficient skills and experience can have fatal consequences, as evidenced by the 15 deaths in Rudraprayag in June 2024.

While the commission's decision is understandable, the process for acquiring this `Additional Endorsement to Driving License' in many states is problematic. Often, the endorsement is obtained purely through an online process, without any practical assessment or testing and merely involves filling out forms and attesting to certain statements [9]. The lack of a meaningful evaluation raises questions about the actual value of this accreditation, undermining its intended purpose.

While Indian transport law may be permissive with such cost-cutting techniques, the laws of physics operate with an iron fist. This regulatory leniency results in frequent accidents, leading to road blockages, delays in the delivery of essential supplies, and elevated risks to both drivers and communities [10]. These issues underscore the urgent need for a robust and innovative solution to enhance transportation safety in these regions. 

The objectives of this project are focused on the development of an advanced device designed to monitor and detect overloaded and inadequately secured loads on commercial vehicles. The proposed device utilises state-of-the-art load sensors to provide accurate, real-time data on the weight distribution and security of cargo. By delivering timely alerts about unsafe load conditions, the device aims to significantly enhance safety on mountainous roads. Additionally, the device is designed to improve logistical efficiency by minimising vehicle breakdowns and delays associated with overloading. It also supports regulatory compliance by offering a practical tool for enforcing safe loading practices. The anticipated benefits include a reduction in long-term operational costs for transport companies through decreased vehicle wear and tear and avoidance of accident-related expenses. Furthermore, by optimising vehicle operations and reducing fuel consumption, the system contributes to environmental sustainability, aligning with broader goals of reducing emissions and promoting efficient transport practices.

\section{Literature Review}

Regulating traffic flow without hindering it, and managing the logistics chain without disrupting its efficiency, are two perennial challenges that successive governments have endeavoured to address. These issues become particularly significant when discussing legislation aimed at controlling the load on commercial vehicles. Two key socioeconomic questions arise from such regulatory efforts:

\vspace{0.5\baselineskip}

1. Impact on Traffic Flow Due to Load Estimation: A primary concern is whether the process of load estimation will lead to significant traffic delays. Requiring truck operators to divert from established routes to obtain clearance at designated locations could result in substantial disruptions. This scenario is especially critical for perishable goods, where time is of the essence. The economic implications of delays further exacerbate this issue, as the rising costs of fuel and the need for timely deliveries are paramount for logistics firms. Moreover, imposing additional steps for load verification could lead to inefficiencies that ripple through the supply chain, ultimately affecting consumers. Therefore, it is essential that any load verification processes are integrated into the existing traffic flow to avoid unnecessary disruptions, particularly as logistical costs continue to escalate. 

2. Potential Backlash from the Trucking Industry: 
Increased regulation may provoke backlash from the trucking sector, which is predominantly composed of blue-collar workers who often feel disenfranchised by governmental oversight. This concern is underscored by the military adage attributed to General Omar Bradley: “Amateurs talk strategy; professionals talk logistics” [11]. The United States' global dominance is, in part, a function of its exceptional logistical capabilities, which enable military force projection worldwide within 24 hours. Similarly, the U.S. industrial sector benefits from a robust logistics framework, reliant on the contributions of truckers who facilitate the movement of goods. India's trucking industry, by comparison, is comprised of a rather obsolete, outdated vehicles and little-to-no consolidation. While nationwide delivery services like Delhivery or DTDC operate with astounding efficiency, it is exponentially more difficult to transport tonnes of fruit than a single package. Transportation in India is generally run by localized businesses, who seek to establish a monopoly under their area of operations. New regulatory burdens could jeopardise this equilibrium and lead to a potential labour shortage if truckers perceive these regulations as excessively punitive. 

\vspace{0.5\baselineskip}

Furthermore, many truck drivers already face significant financial pressures, including rising operational costs and the burden of taxes. The addition of stringent regulations may exacerbate their grievances, leading to heightened resistance against governmental oversight. Should profit margins dwindle due to these regulatory measures, there exists a tangible risk that many truckers may exit the industry, thus undermining the critical supply chain that supports economic activity.

Currently, most governments impose standardised limits on the maximum permissible weight for commercial vehicles and specific weight restrictions per axle. However, these regulations often fail to account for essential factors such as road conditions, topography, and incline variations. The enforcement of these regulations poses its own challenges, as trucks are not routinely subjected to inspections, which allows many logistics companies to overload their vehicles. The potential revenue generated from overloading frequently outweighs the penalties imposed for non-compliance, resulting in a regulatory environment that struggles to maintain efficacy.

One potential solution to this challenge is the implementation of weighbridges, strategically located at critical checkpoints such as border crossings, bridges, and toll booths to measure a vehicle’s total weight in real time. This system offers a non-intrusive means of ensuring compliance without necessitating detours for trucks. 

\subsection{An overview of weighbridges}

Bwire and Nairobi [12] conducted an extensive analysis of the implementation and regulation of weighbridges in Kenya, highlighting how the Traffic Act of Kenya Cap 403 (2018) defines permissible gross vehicle weights (GVW) by axle configurations. For example, the 2 Configuration (two axles with single wheels) permits up to 18,000 kg, while the 2A Configuration (two axles with front single and rear double wheels) also allows up to 18,000 kg. Larger configurations, like the 6A and 7 Configurations for articulated trucks, can support up to 56,000 kg for vehicles with seven axles. 

Fixed weighbridges are permanent structures installed along roads, offering high accuracy and durability, making them ideal for high-traffic areas where enforcing load regulations is critical. In contrast, portable weighbridges are mobile units that can be installed temporarily for unscheduled checks. While they offer flexibility, they are generally less durable and less accurate, often deployed in remote areas where frequent location changes are necessary.

Due to their high accuracy, fixed weighbridges are the most commonly used in Kenya. They can measure both gross vehicle weights (GVW) and axle loads, enabling trucks to redistribute loads when overloaded. Smaller single axle weighbridges, which weigh one axle at a time, are easier to relocate but slower, as operators must sum individual axle readings to calculate the total weight. As noted by Victor [13], this process can be time-consuming. On the other hand, axle unit weighbridges weigh multiple axles in one operation, providing faster and more accurate results. These are commonly used in high-traffic areas, with platform sizes ranging from 3.2m x 3m to 3.2m x 4m [12],[13].

Multi-deck weighbridges, which use multiple platforms to weigh multi-axle vehicles simultaneously, offer the highest levels of accuracy and efficiency, making them the preferred choice for monitoring trucks on major highways. Mobile weighbridges consist of movable pads placed on the road surface, where axle weights are calculated by summing the wheel loads. These weighbridges are typically used for random, unscheduled checks, but they require levelling mats and frequent calibration to ensure accuracy.

Weighbridges operate in two main modes: static and dynamic. In the static method, vehicles must stop on the weighbridge platform, offering higher accuracy, which is essential for legal enforcement. However, it is time-consuming, leading to potential delays in high-traffic areas. In this mode, the total mass transmitted from all axles to the wheels is measured over a 15-second time span [14]. The dynamic method, or Weigh-in-Motion (WIM), allows vehicles to be weighed while moving, reducing traffic congestion. However, while WIM systems help maintain traffic flow, they are less accurate than static methods, making them less suitable for legal proceedings like establishing liability in case of accidents [15].

Accuracy is critical for prosecuting truck operators for overloading. In Kenya, permissible errors depend on the weighbridge's capacity, ranging from 20 kg for an 80-tonne capacity weighbridge to 80 kg for a 400-tonne capacity unit upon re-verification. For first-time verification, stricter tolerances apply, such as 10 kg for an 80-tonne weighbridge and 40 kg for a 400-tonne unit. For comparison, New Zealand's tolerances range from ±40 kg for loads between 10 and 40 tonnes, while the US National Institute of Standards Handbook 44 specifies an acceptance tolerance of 0.1\%, equating to $\pm$40 kg for a 40-tonne load [16].

The type of weighbridge selected also depends on traffic volumes and road classification. Multi-deck weighbridges are used on high-traffic roads, while single-axle or portable weighbridges may suffice for roads with lower traffic. Weighbridge installations are capital-intensive projects, with costs influenced by platform size, load capacity, and supporting infrastructure. In 2008, single-axle weighbridges cost between 0.4 and 1 million USD, while multi-deck weighbridges ranged from 6.0 to 8.0 million USD. Adjusted for inflation, these costs would be approximately \$584,198 to \$1,460,495 for single-axle weighbridges and \$8,762,971 to \$11,683,961 for multi-deck weighbridges, inclusive of procurement, installation, and maintenance [17].

Despite the high costs, implementation has proven to be fruitful. For instance, in 1995, an Indian government official visiting Chicago observed how technology was integrated to streamline dumping ground procedures. Upon returning to Calcutta, he faced opposition from local mafias but successfully installed a static weighbridge system at a cost of 80 lakhs (then), replacing a trip-based payment system with one based on the tonnage of waste dumped. This reduced the number of trips to 500 and significantly improved the efficiency of truck loading [18]. In more recent times, Mkhize and De Beer [19] conducted a statistical analysis of vehicle loads using a novel Stress-in-Motion (SIM) mechanism at a traffic signal near Heidelberg, Germany. Their study demonstrated strong repeatability and reproducibility, with the SIM system underestimating the actual Gross Vehicle Mass (GVM) by only 6\%. This slight underestimation highlights the system's potential for real-time vehicle load monitoring, complementing existing weighbridge methods and offering a promising alternative for dynamic load assessment in traffic-heavy environments.

\subsection{Alternative approaches}

Jeuken [20] recognised that weighbridges were permanent, capital-intensive investments often unnecessary for simple applications, such as weighing livestock or agricultural produce. To address this issue, he developed a `free-hanging cattle cage' as a low-cost, easily portable alternative to traditional weighbridges. In another project, he aimed to replicate the functionality of a weighbridge within the vehicle by utilising cargo compartments, or bulk tanks, as load receptors in various configurations. He integrated the cargo compartment rigidly with the chassis of a lorry through multiple strain gauge load cells, effectively making the weighing system an integral part of the vehicle. The load cells were strategically positioned between the subframe and the cargo compartment, designed to be robust enough to withstand operational stresses. However, the specific dimensions of these load cells resulted in limited measurement resolution, meaning that accurate weighing was primarily feasible when the vehicle was on a level surface.

Additionally, the project explored a specialised application of this weighing system within semi-trailers. In this configuration, load cells were installed between the tank and the running gear on one side, and between the semi-trailer coupling tray and the truck's subframe on the other. This design enhanced the overall measurement capability while accommodating the dynamic characteristics of trailer operation.

In another project, Longo et al. [21] proposed a simplified physics-based model for a compact angular head, known as RHEvo, which was primarily used for hemming in production lines. This tool consisted of mechanical components, including springs, rollers, skates, and bearings. After developing the model, the authors employed a neural network to estimate the current state of these internal components. Their physical analysis showed that aging impacted the elastic coefficient of the springs due to fatigue degradation. They also calculated the remaining useful life (RUL) of the internal springs using a stochastic model. This approach was applicable to various devices that utilised springs, including automobile suspension systems, weapon recoil mechanisms, and engine shut-off valves.

Tinga and Loendersloot [22] also discussed methods and tools that enhanced predictive maintenance. The authors highlighted that vibration-based machinery health monitoring techniques were effective for detecting damage, diagnosing system health, and predicting the remaining life of machinery. They introduced a decision support tool that guided users in selecting the most suitable predictive maintenance strategies and monitoring techniques.

Building on this, Yang et al. [23] applied neural network algorithms trained on engine data to predict vehicle overloading. Their model aimed to detect engine performance patterns correlated with overload conditions, providing a proactive approach to fleet management.

Similarly, Praveena et al. [24] developed a real-time load monitoring system designed to ensure trucks comply with weight regulations. This system used load cells installed between the chassis and trailer to continuously measure weight. When the permissible weight limit was exceeded, the system automatically disconnected the battery, preventing the engine from starting. To address uneven load distribution on inclines, a gyroscopic sensor was integrated, allowing the system to disable this restriction when the vehicle was on a slope.

Complementing this work, Chen and Chen [25] designed a module that utilised data from a Tire Pressure Monitoring System (TPMS) to detect overloading. By correlating changes in tyre pressure with vehicle weight, their system provided an additional layer of monitoring, ensuring compliance with weight regulations and helping prevent overloading.

Arena et al. [26] conducted a comprehensive survey on predictive maintenance mechanisms, focusing on how these systems could identify overloading patterns in trucks and commercial vehicles. By establishing a baseline and monitoring the frequency of maintenance recommendations, one may be able to infer potential overloading occurrences. This approach offers insights into operational stresses and fleet management inefficiencies. However, the analysis rests on the assumption of ceteris paribus—that all external factors remain constant. It assumes that drivers are adhering to optimal driving practices, no accidents are occurring, road conditions are normal, and vehicle loads are evenly distributed. While this provides a controlled environment for analysis, real-world variables such as driver behaviour, uneven load distribution, mechanical wear, and changing road conditions introduce significant variability, affecting the accuracy of the findings. To address these issues, further research is required to develop adaptive models that consider these dynamic factors in predictive maintenance strategies.

\section{Methods}

\subsection{Components utilized}

A load cell operates on the principle of converting mechanical force into an electrical signal through the use of strain gauges [27]. When an external load is applied to the load cell, it causes a deformation in the load cell’s structure, typically made of a metallic material. This deformation can be in the form of compression, tension, or bending, depending on the design of the load cell. 

Strain gauges, which are thin, flexible electrical resistors, are bonded to the surface of the load cell. As the load induces deformation, the strain gauges also stretch or compress, leading to a change in their electrical resistance [28]. The resistance change is minimal, so it is measured using a Wheatstone bridge circuit, which amplifies the small changes in voltage that correspond to the deformation. This circuit consists of four resistive elements, where the load cell itself is typically one of the resistors. The output voltage from the Wheatstone bridge is directly proportional to the load applied, allowing for precise measurements [29].

To ensure the accuracy of the load cell, it must undergo a calibration process. During calibration, known weights are applied to the load cell, and the output voltage is recorded to create a reference scale. This process allows the system to account for any variations due to temperature, material properties, or manufacturing tolerances. Once calibrated, the load cell can provide real-time feedback on the weight applied, making it suitable for various applications, including monitoring loads on vehicles, measuring forces in industrial settings, and ensuring compliance with safety regulations [30].

To accurately convert this signal for further processing, the HX711 is employed. This precision 24-bit analogue-to-digital converter (ADC) is specifically designed for weigh scales, providing high-resolution measurements while simplifying the interface between the load cell and a microcontroller [31]. The HX711 incorporates built-in gain settings, allowing for easy amplification of the small voltage changes from the load cell [32]. Its low power consumption and compact design make it ideal for portable applications. By using the HX711, the project achieved accurate and stable weight measurements with minimal complexity.

The microcontroller used here was the Arduino Uno. It features a straightforward design that includes a microcontroller, a variety of digital and analog input/output pins, and built-in communication interfaces [33]. Its compatibility with a vast array of sensors, actuators, and modules allowed us to easily scale up between prototypes. The extensive community support and resources available for the Arduino Uno, including libraries and tutorials, further enhance its accessibility for those looking to develop innovative solutions [34].

\subsection{Prototype 1}

\begin{figure}[H]
    \centering
    \includegraphics[width=\columnwidth, height=10cm, keepaspectratio]{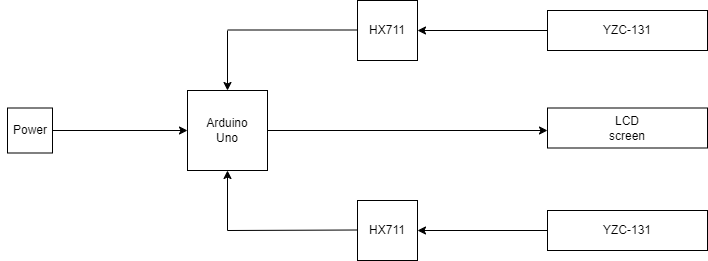}
    \caption{Simplifed Block Diagram} 
    \label{fig:photo} 
\end{figure}


The centre of gravity (CoG), which this project estimates, is the point where the total weight of an object is considered to be concentrated, serving as a critical parameter for evaluating stability and balance, particularly in dynamic systems such as vehicles. To achieve accurate assessment of load distribution, Prototype 1 employed two load cells positioned to measure the weight on the left and right sides of the vehicle, respectively. By combining these measurements with the known breadth of the vehicle, it calculated the lateral position of the CoG.

This information is essential not only for determining whether the vehicle is overloaded but also for evaluating how securely the load has been fastened. If the calculated CoG is significantly offset from the centreline, it signals an uneven load distribution. In scenarios such as navigating sharp turns on mountainous roads, an improperly distributed load could result in increased inertia in one direction, raising the risk of vehicle rollover due to an imbalanced force during cornering. Therefore, this lateral CoG calculation enhances vehicle safety by providing insight into both the overall load status and its spatial distribution.
\begin{figure}[H]
    \centering
    \includegraphics[width=1\linewidth]{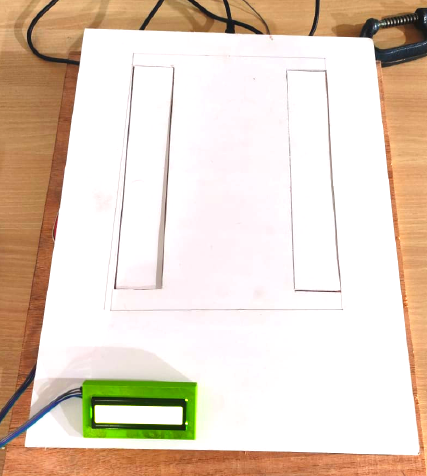}
    \caption{Prototype 1 (polished)}
    \label{fig:enter-label}
\end{figure}

Let \( W_R \) be the reading from the right load cell and \( W_L \) be the reading from the left load cell. The total weight of the vehicle \( W_{\text{total}} \) is the sum of both readings:

\[
W_{\text{total}} = W_R + W_L
\]

Then the lateral position of the CoG can be found using the ratio of the weights and the vehicle's breadth \( B \). The distance \( x \) of the CoG from the centreline of the vehicle is calculated by:

\[
x = \frac{W_L - W_R}{W_{\text{total}}} \times \frac{B}{2}
\]

This prototype assumed the vehicle’s breadth, defined as the distance between the two load cells, to be constant. It provided both the centre-of-gravity (CoG) calculation and an assessment of whether the vehicle was within permissible safety limits on the LCD screen. The load cells used had a capacity of 5 kg each, so the threshold for classifying the vehicle as unsafe was set at 9.5 kg.

\begin{figure}[H]
    \centering
    \includegraphics[width=1\linewidth]{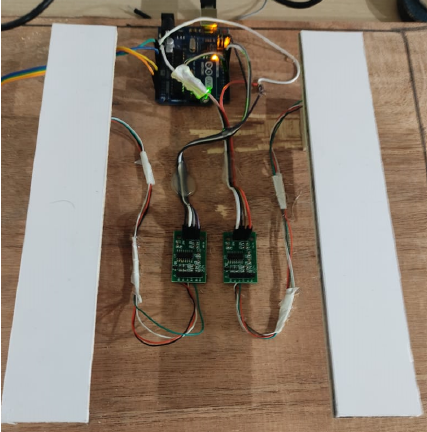}
    \caption{Prototype 1 (bare)}
    \label{fig:enter-label}
\end{figure}

\subsection{Prototype 2}

\begin{figure}[H]
    \centering
    \includegraphics[width=1\linewidth]{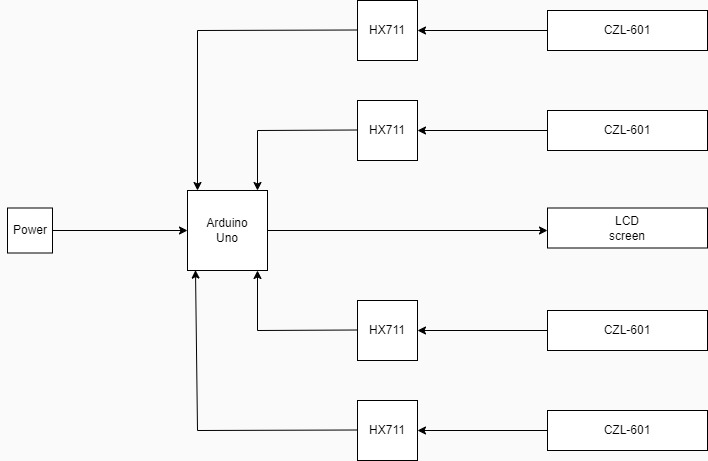}
    \caption{Simplified Block Diagram}
    \label{fig:enter-label}
\end{figure}

This prototype was designed with one load cell positioned under each tyre. This configuration enabled the project to accurately determine the centre of gravity both longitudinally and laterally. Additionally, it facilitated the identification of any load imbalances by dividing the vehicle into four sectors. As this prototype utilized better load cells capable of measuring 120 kilograms, the threshold for alerting the user on the LCD was set to 400 kilograms\\

Calculations:

\begin{itemize}
    \item \( W_1 \) = Front Left (FL) load cell reading.
    \item \( W_2 \) = Front Right (FR) load cell reading.
    \item \( W_3 \) = Rear Left (RL) load cell reading.
    \item \( W_4 \) = Rear Right (RR) load cell reading.

    \item \( L \) = Distance between the front and rear axles (wheelbase).
    \item \( T \) = Distance between the left and right tyres (track width).
\end{itemize}

The total weight of the vehicle is:
\[
W_{\text{total}} = W_1 + W_2 + W_3 + W_4
\]

The longitudinal position of the CoG is calculated relative to the front axle. It can be determined using the following equation:

\[
X_{\text{cg}} = \frac{(W_3 + W_4) \cdot L}{W_{\text{total}}}
\]

The lateral position of the CoG is calculated relative to the vehicle's centreline using the following equation:

\[
Y_{\text{cg}} = \frac{(W_1 + W_3) \cdot T}{W_{\text{total}}}
\]\\

\hspace{-0.5cm} The front weight of the vehicle:\[W_{\text{front}} = W_1 + W_2\]
     The rear weight of the vehicle:\[
W_{\text{rear}} = W_3 + W_4
\]
     The left weight of the vehicle:\[
W_{\text{left}} = W_1 + W_3
\]
     The right weight of the vehicle:\[
W_{\text{right}} = W_2 + W_4
\]

If \( W_{\text{front}} > W_{\text{rear}} \), the vehicle is front-heavy. If \( W_{\text{rear}} > W_{\text{front}} \), the vehicle is rear-heavy.

If \( W_{\text{left}} > W_{\text{right}} \), the vehicle is left-heavy. If \( W_{\text{right}} > W_{\text{left}} \), the vehicle is right-heavy.\\

Percentage of total weight in each quadrant:

\[
\text{Front Left (FL) Percentage} = \frac{W_1}{W_{\text{total}}} \times 100
\]

\[
\text{Front Right (FR) Percentage} = \frac{W_2}{W_{\text{total}}} \times 100
\]

\[
\text{Rear Left (RL) Percentage} = \frac{W_3}{W_{\text{total}}} \times 100
\]

\[
\text{Rear Right (RR) Percentage} = \frac{W_4}{W_{\text{total}}} \times 100
\]

Higher weight percentages in a certain quadrant (set to 30\%) will indicate a weight imbalance in that sector on the LCD.

\begin{figure}[H]
    \centering
    \includegraphics[width=1\linewidth]{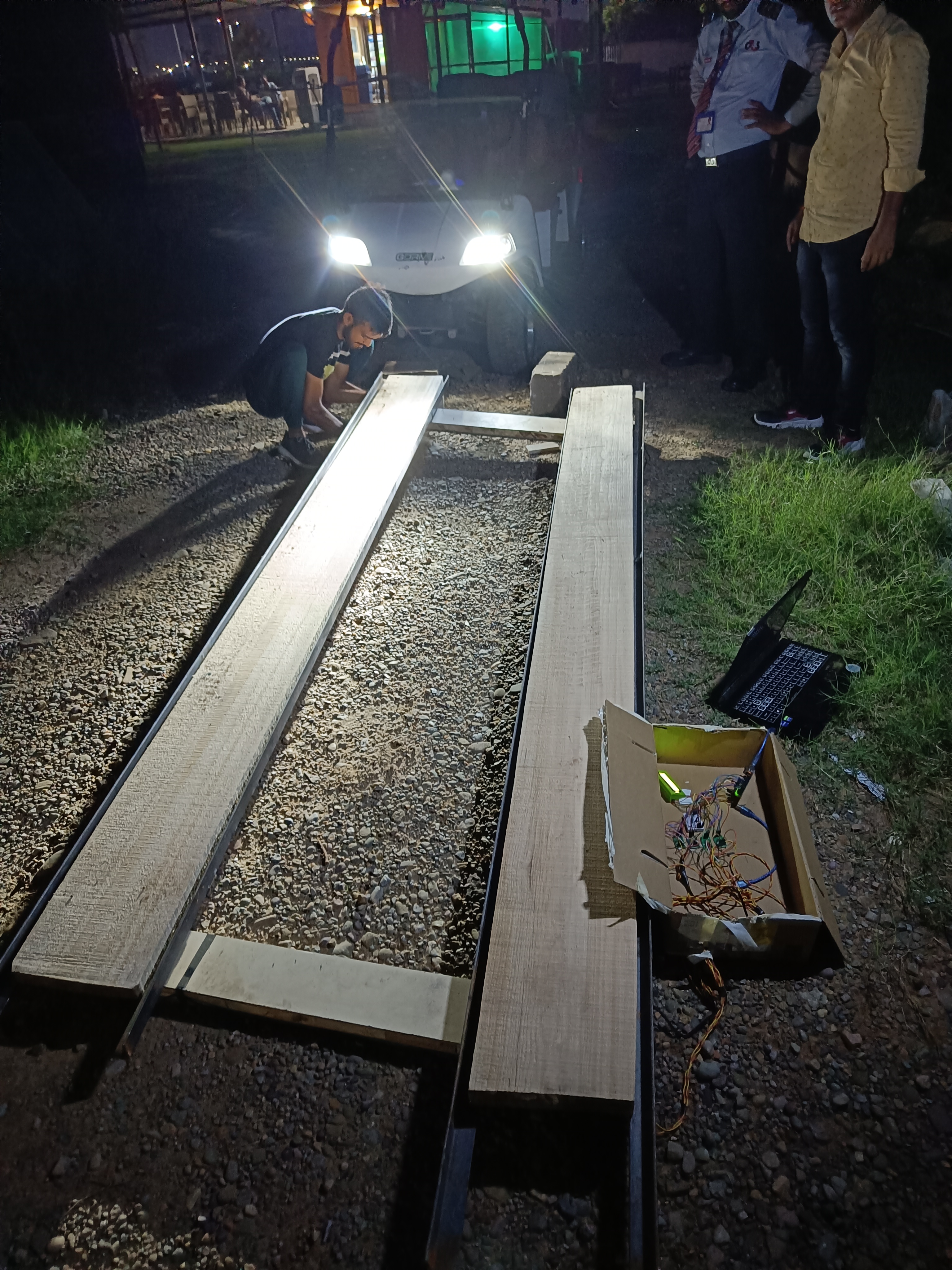}
    \caption{Testing of Prototype 2}
    \label{fig:enter-label}
\end{figure}

\section{Results and Future Research}

The electric golf cart underwent testing with various weights strategically placed in different locations to evaluate the functionality of the prototype. The results indicated no latency or substantial errors. Although a mathematical quantification of error has not been conducted, extensive testing of the system over several hours has consistently demonstrated no occurrences of false positives (i.e., detecting weight when none is present) or false negatives (i.e., failing to detect weight when it is present). Standard dumbbells were employed for this testing, leading to the successful completion of the second prototype, which demonstrated the capability to detect weight overload and identify weight imbalances. While core functionalities have been achieved, additional research is warranted to further enhance the system.

The authors intend to develop a variable system to accommodate vehicles with diverse wheelbases. This may be implemented through a drawer-slide mechanism, facilitating expansion or contraction as necessary. Alternatively, a discrete fastening system may be utilised, with load cells secured at intervals of 50 cm. This would involve drilling holes into the underlying platform at specified intervals to provide adequate support for the load cells. Similar modifications would be required for the track length.

For subsequent iterations, it is imperative to incorporate load cells with higher capacity ratings. The current testing has been conducted using an electric golf cart, which poses minimal challenges; however, when scaling to larger vehicles, such as trucks, load cells will need to measure in tonnes. Furthermore, materials with enhanced tensile strength must be utilised to accommodate the expected weight loads. Simulations will also be conducted to evaluate designs capable of supporting loads of at-least 5 tonnes in total.

The existing prototype features a modular design that allows for disassembly and reassembly within approximately 10 to 15 minutes. Ensuring that the system can handle these weight capacities and accommodate various vehicles could facilitate its deployment to assist trucking companies in mitigating accidents.

It is essential to address the underlying issue of lax regulatory standards. Remediating this root cause will necessitate collaboration between governmental bodies and law enforcement to enact stricter legislation regarding vehicle loading and safety protocols.

Additionally, the authors are contemplating the integration of supplementary sensors, such as temperature sensors, to account for factors such as thermal expansion that may influence the system's performance. This enhancement would enable a more comprehensive assessment and ensure the reliability and safety of the load detection mechanism.

\section{Conclusion}

While weighbridges provide a time-tested solution for verifying vehicle weight compliance, their high cost—approximately 1.5 million dollars—and static nature pose significant challenges. Once installed, relocating them is complex, and their inaccuracy in weigh-in-motion methods has led some legislatures to mandate repeated re-verification to ensure the reliability of readings for determining truck operator liability. This project addresses these issues and has already demonstrated effective operation with an electric golf cart. With additional modifications and considerations for variables such as thermal expansion and pressure, the system has the potential to become a comprehensive solution applicable in various contexts, including hilly regions and operational sites such as docks and border crossings to detect illicit materials. The proposed system is low-cost, portable, and has not yet produced any erroneous readings.

\section*{Acknowledgement}

The first iteration of this project was honoured with the prestigious SP Dutt Innovation Award, securing first prize and a grant of INR 25,000. Additionally, the project received further support through the esteemed Plaksha Summer Internship Programme, where its second iteration was awarded the Silver Medal.

\end{document}